# Magnetic impurity formation in quantum point contacts

Tomaž Rejec[1] & Yigal Meir[1,2]

[1]*Department of Physics, Ben Gurion University, Beer Sheva 84105, Israel*

[2]*The Ilse Katz Center for Meso- and Nano-Scale Science and Technology, Ben Gurion University, Beer Sheva 84105, Israel*

**A quantum point contact (QPC), a narrow region separating two wider electron reservoirs, is the standard building block of sub-micron devices, such as quantum dots - small boxes of electrons, and qubits - the proposed basic elements of quantum computers. As a function of its width, the conductance through a QPC changes in integer steps of $G_0 = 2e^2/h$, signalling the quantization of its tranverse modes.[1,2] Such measurements also reveal an additional shoulder at a value around $0.7 \times G_0$,[1-4] an observation which remains a puzzle even after more than a decade. Recently it has been suggested[5,6] that this phenomenon can be explained if one invokes the existence of a magnetic impurity in the QPC at low densities. Here we present extensive numerical density-functional calculations that reveal the formation of a magnetic moment in the channel as the density increases above pinch-off, under very general conditions. In addition we show that such an impurity will also form at large magnetic fields, for a specific value of the field (corresponding to a degeneracy point between the upper spin state in the first mode and the lower spin state in the second mode), and sometimes even at the opening of the second mode in the QPC. Beyond explaining the source of the "0.7 anomaly", these results may have far reaching implications on spin filling of electronic states in quantum dots and on dephasing of quantum information stored in semiconductor qubits.**

A quantum point contact is usually formed by applying a negative voltage to a split-gate (Fig. 1a), depleting the electrons in the two-dimensional electron gas (2DEG) under it to form a narrow and short constriction connecting large, two-dimensional regions of 2DEG. As a function of the applied gate voltage the number of occupied quantised transverse modes can be changed. As each mode contributes $G_0$ to the conductance (due to spin degeneracy), the conductance rises in steps quantized at integer multiplies of $G_0$.[1,2] A magnetic field lifts the spin degeneracy, leading to steps in multiples of $e^2/h$. Surprisingly, many experiments observe, at zero magnetic field, an additional shoulder near $0.7 \times G_0$, a feature usually referred to as the 0.7 anomaly,[3,4] which merges smoothly with the $0.5 \times G_0$ plateau in a large magnetic field. Although less robust, an analogous anomaly was observed in the transition to the second conductance plateau at about $1.7 \times G_0$.[4] A similar conductance structure was also found at large magnetic fields near crossings of spin-up and spin-down modes of different subbands.[7]

There have been several attempts to explain the 0.7 anomaly in terms of an antiferromagnetic Wigner crystal[8], spontaneous subband splitting,[9,10,11] or by assuming that the QPC supports a local quasi-bound state.[6] In the latter case, as one electron is transported through this state, Coulomb interactions suppress the transport of an electron with opposite spin, reducing the conductance to around $0.5 \times G_0$. The Kondo effect, the screening of this local spin by the conduction electrons, enhances the conductance at low temperatures toward $G_0$. Indeed, experiments[5] reveal features characteristics of the Kondo effect and the continuous evolution of the conductance from $0.7 \times G_0$ at higher temperatures to $G_0$ at low temperatures. But how can a QPC, being an open system, support a quasi-bound state? Previous numerical investigations report conflicting results.[12,13,14] Below we present detailed spin-density-functional theory[15] (SDFT) calculations showing that such a local moment can indeed form as the conductance of a QPC rises towards the first plateau. Additionally we present evidence that the formation of a quasi-bound state may also lead to the observed 1.7 anomaly and the anomaly at the crossing of subbands in large magnetic fields.

The set-up we used in our calculation is shown in Fig. 1b. In order to make the calculation tractable we modelled the reservoirs as semi-infinite quantum wires. They are wide enough to carry many modes and thus resemble well the two-dimensional reservoirs in experimental set-ups. Using SDFT within the local spin-density approximation (LSDA) we calculated the spin-densities of the 2DEG and the charge distribution on the electrodes self-consistently. The nonlinearity of equations may lead to several stable solutions that differ in their energy (details of our numerical approach are given in the Methods section below). Here we present results for a specific QPC, as described in Fig. 2. However, the results are generic: we studied QPCs of lithographic length from 100 nm to 400 nm, of lithographic width from 150 nm to 250 nm, and electron densities from $10^{11}$ to $2\times10^{11}$ cm$^{-2}$, with very similar results.

We first consider a QPC in the absence of a magnetic field. For gate voltages corresponding to conductance plateaus there is a unique solution to our equations, which shows no spin polarisation. Although such a solution is also present in the regions between the plateaus, additional solutions exhibiting spin polarisation in the QPC appear there. We classify these solutions according to their spatial symmetry: in the "symmetric" and "antisymmetric" solutions the polarisation is the same or opposite on the two sides of the QPC, respectively. In Fig. 2a the energies of the polarised solutions relative to the energy of the unpolarised solution are plotted for gate voltages from pinch-off to the second conductance plateau. The symmetric solution, when present, is always the ground state of the system.

Fig. 2b shows the evolution of spin densities of the two polarised solutions from pinch-off to the point where a spin 1/2 magnetic moment forms in the QPC. At point A the QPC is pinched-off: there is an extended region about the centre of the QPC where the density vanishes as the potential is much higher than the Fermi energy. The potential decreases towards the reservoirs and at some point it crosses the Fermi energy. Here are regions where the electron gas is polarised: the density is very low and the gain in (negative) exchange energy overweights the additional kinetic energy incurred by polarisation. The two polarised regions on both sides of the QPC do not overlap in this regime. The degeneracy of the ground state is thus fourfold: each of the polarised regions is spin-degenerate. With increasing gate voltage the potential at the QPC gets lower and the electron density drifts inwards, forming narrow fingers which eventually reach the centre of the QPC (point B in Fig. 2). In the process the polarised regions become increasingly decoupled from the reservoirs: two quasi-bound states, each corresponding to roughly one electron, form on each side of the QPC. As the potential barrier at the centre of the QPC gets weaker, the tunnelling probability through the barrier becomes appreciable and the conductance increases from zero. On increasing the gate voltage even further, a different configuration becomes energetically favourable in the symmetric subspace: a single electron forms a weakly coupled quasi-bound state at the centre of the QPC, accompanied by two electrons of opposite spin in regions further away towards the reservoirs (point C in Fig. 2). The degeneracy of the ground state is now twofold: the QPC acts as a spin 1/2 magnetic impurity. The spin-resolved local density of states (LDOS), i.e. the number of states per energy and length interval, shown in Fig. 3, provides additional insight into this state. The QPC potential for one of the spin components assumes a double-barrier form (due to Friedel oscillations), and a resonant state forms in its minimum.

The length of the QPC affects the formation of the spin 1/2 magnetic moment. In very short contacts the transition to a well defined quasi-bound state does not take place at all: as the two polarised regions merge at the centre of the QPC, the conductance already reached the first plateau. For longer contacts the transition to the magnetic moment state shifts towards the pinch-off. In very long contacts, the configuration in Fig. 2b (symmetric solution at point C) evolves into an antiferromagnetically ordered chain.moment state shifts towards the pinch-off. In very long contacts, the configuration in Fig. 2b (symmetric solution at point C) evolves into an antiferromagnetically ordered chain.

The situation is somewhat similar on the rise from the first to the second conductance plateau: the density of electrons in the second mode is low there and exchange again stabilizes states with a polarised QPC (inset to Fig. 2a). This mechanism is not as efficient here as it was in the first mode: polarisation in the second mode induces, due to exchange, also partial polarisation in the first mode. As the density in the first mode is large, there is a high kinetic energy cost involved. This is consistent with the experimental observation that not all QPCs exhibit the 1.7 anomaly.

In an external in-plane magnetic field the energies of transverse modes for the two spin components split. The resulting polarised non-degenerate solution does not generally support a quasi-bound state. However, as shown in Fig. 4a, at a particular value of the field the energies of spin-up electrons (those with spin parallel to the field) in the first mode and of spin-down electrons in the second mode cross. By tuning the gate voltage one can also make the energy of the degeneracy point coincide with the Fermi energy. Fig. 4b shows the evolution of spin-densities with magnetic field in vicinity of this crossing. At the degeneracy point a well-defined quasi-bound state forms, but, unlike the zero field solution, with a definite spin, determined by the field. This may be the source of the "0.7 analogs" observed at high magnetic fields.[7]

The formation of a local spin-degenerate quasi bound state, supported by the extensive SDFT calculations presented here, is a necessary condition for the Kondo effect, which is beyond the LSDA employed here. Interestingly the calculation indicates that as the QPC opens up, first two such states form on the two sides of the QPC. This may lead to the physics of the two-impurity Kondo model. Depending on the ratio of the coupling between these impurities, and their couplings to their respective reservoirs, one would expect to observe a zero-bias anomaly, with split zero-bias peak, in this regime.[16,17] The splitting should increase with increasing conductance. The formation of such polarised state at the QPC may also affect the spin-filling of quantum dots, which are formed by two QPCs. Additionally, as quantum dots have been proposed as qubits – the building blocks of quantum computers, these degenerate quasi-bound state must be considered seriously – the degeneracy allows decoherence of quantum processes at very low temperatures. Since short decoherence time will deteriorate the performance of any quantum computer, one must make sure that the QPCs forming the qubits are outside the quasi-bound state formation regime.

**Methods**

We treated the 2DEG, electrodes and the donor layer as a set of three electrostatically coupled strictly two-dimensional systems. We assumed the donor layer was uniform and fully ionized. Then, according to spin-density-functional theory[18] the properties of the system can be uniquely determined in terms of spin-densities $n_\uparrow(\mathbf{r})$ and $n_\downarrow(\mathbf{r})$ of the 2DEG and by the distribution of charge on electrodes $n_{el}(\mathbf{r})$. In particular, the energy of the system is a functional of the densities:

$$E[n_\uparrow, n_\downarrow, n_{el}] = E_{es}[n_\uparrow, n_\downarrow, n_{el}] + T_s[n_\uparrow, n_\downarrow] + E_{xc}[n_\uparrow, n_\downarrow] + $$
$$+ \frac{1}{2} g\mu_B \int d\mathbf{r} B(\mathbf{r})[n_\uparrow(\mathbf{r}) - n_\downarrow(\mathbf{r})] - \sum_i N_i V_i.$$

Here $E_{es}[n_\uparrow, n_\downarrow, n_{el}]$ is the electrostatic energy of the charge distribution. The presence of the semiconductor was taken into account through the use of the GaAs dielectic constant $\kappa = 12.9$ and a modified form of the Coulomb interaction, reflecting the dielectric mismatch at the surface.[19] We treated the 2DEG quantum-mechanically by including its kinetic ($T_s$) and exchange-correlation ($E_{xc}$) energies into the energy functional, taking into account that GaAs effective mass is 0.067 the bare electron mass.

We used the local spin-density approximation for the exchange-correlation functional, as parametrized by Tanatar and Ceperley[20]. The fourth term in the energy functional is the Zeeman energy due to an in-plane magnetic field, with $g = 1.9$ for GaAs quantum wires.[7] Finally, as we compared the energies of different solutions at fixed voltages between electrodes and the 2DEG, we applied a Legendre transform to the energy functional (the last term in the expression above), with $N_i$ and $V_i$ being the number of electrons and the voltage on $i$-th electrode, respectively, and the sum running over all the electrodes in the system. The correct densities minimize the above functional subject to the applied voltages, and a constraint that the total number of electrons should match the charge provided by the donor layer. The minimization procedure yields Kohn-Sham equations[15] for the 2DEG, with a constant electrostatic potential on each of the electrodes. In each iteration of the self-consistency loop we first solved for the Kohn-Sham scattering states and calculated the density of the 2DEG. Using an iterative approach, we then redistributed the remaining electrons on electrodes in such a way that the potential there assumed the required form. In this step we performed the calculation on a large rectangular box, with periodic boundary conditions, which enabled us to employ the fast Fourier transform method and thus make the calculation efficient. We used the resulting charge distribution to calculate an improved electrostatic potential in the 2DEG. To obtain spin-polarized solutions at zero external magnetic field we broke the symmetry by applying a magnetic field of an appropriate form (spatially symmetric or antisymmetric) in the initial iterations of the self-consistent procedure.

**Acknowledgements**

We thank N. Argaman, R. Baer, D. Goldhaber-Gordon, B.I. Halperin, K. Kikoin and M. Stopa for discussion. This work was supported by the Binational Science Foundation.

**Author Contribution**

Y.M. initiated the project. T.R. performed numerical simulations. T.R. and Y.M. analysed results and co-wrote the paper.

**Author Information**

The authors declared no competing financial interests. Correspondence and requests for materials should be addressed to T.R. (tomaz.rejec@ijs.si).

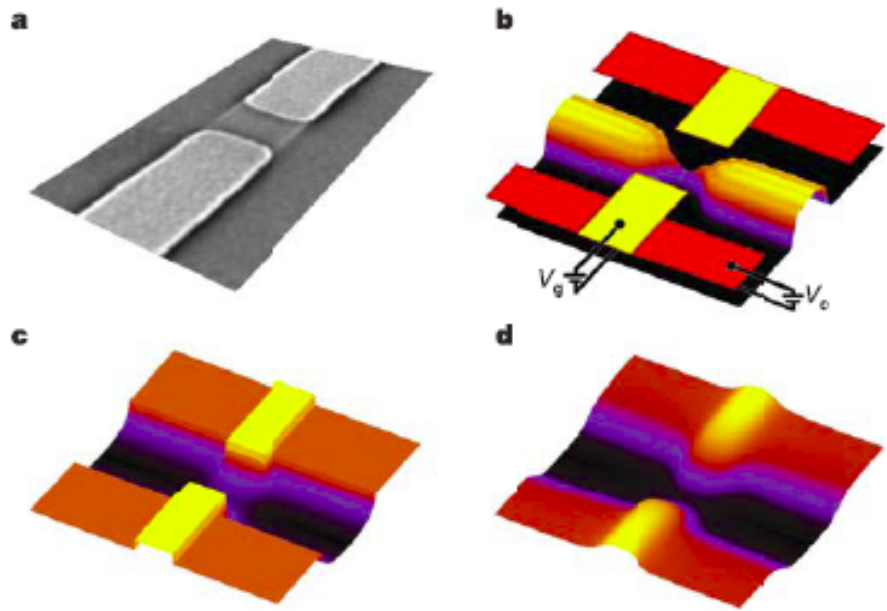

**Figure 1 | Quantum point contact. a**, Micrograph of a split-gate forming a QPC.[5] **b**, The set-up used in the calculation. Voltage ($V_g$) applied to split-gates (yellow) forms a QPC between two quantum wires defined by negatively biased ($V_c$) confining electrodes (red). Also shown is the density of the 2DEG near pinch-off where only one mode is occupied within the QPC, while the quantum wires carry several modes. **c**, **d**, A typical potential in the electrodes plane and in the 2DEG, respectively. Potentials on the electrodes are constant and are the input to the calculation, determining the effective potential in the plane of the 2DEG.

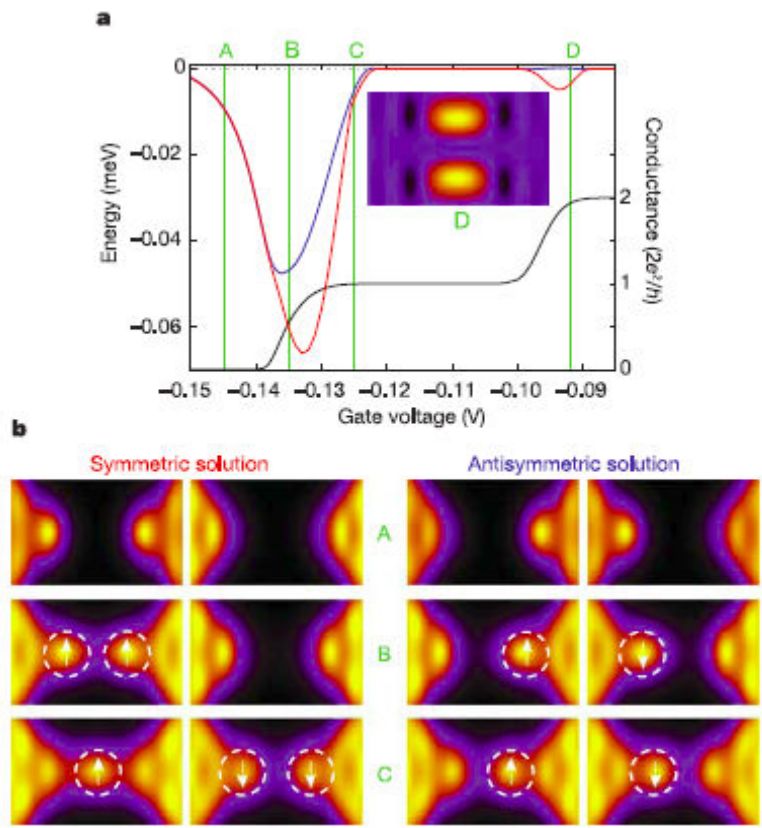

**Figure 2 | States with spin polarisation in a QPC.** The QPC with a lithographic width and length of 200 nm and 250 nm, respectively forms a constriction in a quantum wire with a lithographic width of 300 nm. The confining electrodes are biased to $V_c$ = -0.08 V. The donor layer provides an electron density of $10^{11}$ cm$^{-2}$ is 20 nm below the surface. The 2DEG forms 50 nm below the surface. **a**, Energies of the symmetric (red) and antisymmetric (blue) spin-polarised solutions relative to the the energy of the unpolarised solution. Polarised solutions appear in the transition from pinch-off to the first plateau and then on the rise to the second plateau. The black line is a rough approximation to the conductance of the QPC, as calculated from the Kohn-Sham wavefunctions of the unpolarised solution (righthand scale). Note, however, that for the lowest energy, polarised solution the first conductance plateau will start around $V_g$ = -0.122 V, on the right of point C. **b**, Spin-densities of the symmetric and antisymmetric solutions for spin-up electrons (left columns) and spin-down electrons (right columns) at three values of gate voltage as indicated in panel a. A 400 nm long and 100 nm wide region about the centre of the QPC is shown. A, The QPC is pinched-off. There are two polarised regions on each side of the QPC. B, The two polarised regions

begin to overlap. C, In the symmetric solution the electrons in the QPC rearrange in such a way that a spin 1/2 magnetic moment forms. The inset to panel a shows spin polarization in the symmetric solution just below the second conductance plateau, at point D.

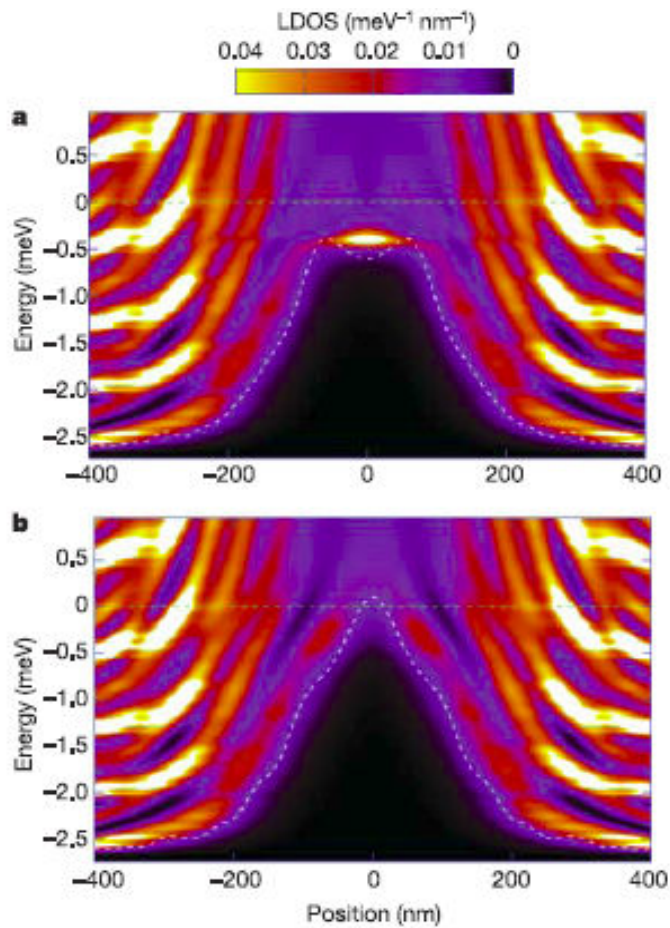

**Figure 3 | Formation of a magnetic moment. a**, Spin-up and **b**, spin-down local densities of states, integrated over the cross-section of the QPC, for the QPC from Fig. 2 at $V_g$ = -0.125 V. The Kohn-Sham potential for electrons in the lowest transverse mode is also shown (white dashed line). The potential for spin-up electrons has a double-barrier form near the centre of the QPC and supports a quasi-bound state about 0.5 meV below the Fermi energy (green dashed line). Spin-down electrons also form quasi-bound states in the shoulders of the potential on both sides of the QPC. The bright stripes on both sides of the QPC correspond to the quasi-one dimensional bands of the reservoirs.

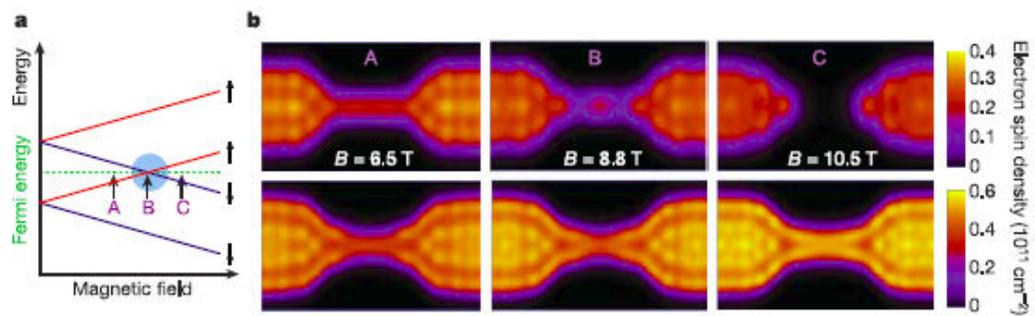

**Figure 4 | Quasi-bound state in a large magnetic field. a**, The Zeeman splitting of mode energies at the top of a QPC. The energy of the spin-up electrons in the lowest mode crosses the energy of the spin-down electrons in the higher mode at some value of the magnetic field. At a specific value of the gate voltage this crossing occurs near the Fermi energy. **b**, Spin-up (top row) and spin-down (bottom row) densities at three values of a magnetic field $B$ near the degeneracy point. A 800 nm long and 250 nm wide area is shown. At $B$ = 6.5 T (point A) one mode is open for both spin orientations. At $B$ = 10.5 T (point C) there are two open modes for spin-down electrons and none for spin-up electrons. In between the bands cross and a quasi-bound state forms in the QPC ($B$ = 8.8 T, point B). The QPC parameters are as in Fig. 2 except for the lithographic length which is 300 nm. The gate voltage is $V_g$ = -0.102 V.